# High Energy electron and proton acceleration by circularly polarized laser pulse from near critical density hydrogen gas target

ASHUTOSH SHARMA*
ELI-ALPS, Szeged, Hungary.

**ABSTRACT**

We demonstrate in this research the quasi-monoenergetic electron and proton acceleration through three dimensional particle-in-cell simulations of short–petawatt circular polarized laser pulse interactions with near critical density hydrogen target. We numerically show that under controlled choice of laser and target parameters, the high energy electrons and protons can be illustrated in experiment at advanced high power laser facilities e.g. ELI-ALPS. We detailed the microphysics involved in the acceleration mechanism, which required investigating the role of plasma density gradients, plasma density, and target thickness. The role of self-generated plasma electric and magnetic fields is depicted on proton energy and density distribution. We numerically investigate here the laser–driven proton acceleration where energetic protons with energies more than 200 MeV and charge in excess of 10 nC[4, 30] and conversion efficiency more than 6% (which implies 2.4 J proton beam out of the 40 J incident laser energy). Additionally and interestingly, we show from simulation study first time the quasi-monoenergetic ring shaped electron beam driven by circularly polarised laser which may prove useful for plasma based-based X-ray source and collimation of positron beam.

*email: ashutosh.sharma@eli-alps.hu

**Introduction**

Ion acceleration with high-intensity lasers has attracted a great deal of attention because accelerated ion beams have extreme laminarity, ultrashort duration and high particle number in MeV energy range. These phenomenal characteristics of ion beam prompted excogitation about a wide range of applications in nuclear and medicine physics[1-7]. The requirement of ion beam with narrow energy spread, high conversion efficiency and compatibility of target with the high-repetition rate laser system, still a challenging task despite a decade plus efforts.

The widely employed schemes for laser driven ion acceleration include target normal sheath acceleration (TNSA)[8-10], radiation pressure acceleration (RPA)[11-19], breakout afterburner (BOA)[20], collisionless shockwave acceleration (CSA)[21-23] and magnetic vortex acceleration (MVA)[24-25]. The basic physics of all the ion acceleration mechanism relies on quasi static electric fields generated due to the charge separation between the lasers accelerated electrons and heavy ions.

Next-generation high power lasers will run at 10 Hz and beyond, bringing applications closer to realisation. However, most research on laser–plasma ion sources has used solid targets, typically metal foils of thickness 1 µm. At high repetition rate, using such targets raises significant challenges with debris, target insertion and unwanted secondary radiation such as bremsstrahlung. Easily replenished, debris-free gaseous targets would alleviate many of these problems. They can generate single-species ion beams, which are challenging for solid targets due to the hydrocarbon layers that typically form on their surface. Thus, the development of gas target at high repetition rate would be transformational in exploring the possibility of an interesting alternative for high quality proton source; since gas targets are relatively less immune to prepulse effect and there is no requirement of pre-shot cleaning as compared to solid target.

Bulanov et. al[26] investigated the petawatt laser pulse interacting with a homogeneous gas plasma, which generates collimated beams of fast ions with energies of several hundred MeV and a relatively small divergence. In this study, the ion acceleration with circularly and linearly polarized laser field was reported through MVA mechanism. Recently MVA has been confirmed experimentally in a shock compressed gas foil[27] where proton acceleration has been reported from a tailored, near solid density gas target. Another recent experiment[28] demonstrated the laser driven ion beams with narrow energy spread and energies up to 18 MeV per nucleon and ~ 5% conversion efficiency (4 J out of the 80 J incident laser energy). The reduction of ion energy spread is reported through the self-generated plasma electric and magnetic fields. There is need of detailed understanding of micro-physics involved in this ion acceleration process which requires to investigate the role of plasma gradient, density and thickness on magnetic field generation and subsequently the peak proton energy. The influence of magnetic field on ion energy spread is point of relevance for applications of ion source.

In our previous research[25], we demonstrated the proton acceleration via MVA mechanism using the PIC simulation. We considered in this case linearly polarised laser (2PW-20fs) interaction with near critical density homogenous hydrogen plasma. Simulation results in this regime show the laser-to-proton conversion efficiency ~1% (for protons with energy > 4 MeV) with peak proton energy ~ GeV, similar to the 2D simulation results of Bulanov et al.[24]. Based on the previous model of MVA ion acceleration, we investigated further in this paper, the circularly polarised laser (2PW-20fs) interaction with inhomogeneous near critical density hydrogen plasma. The extensive study of three-dimensional PIC simulation results show the higher laser-to-proton conversion efficiency ~6% (for protons in energy range of 10-100MeV) with peak proton energy 350 MeV. We demonstrate in the following text, the regime of efficient ion acceleration and shown its dependence on laser-plasma parameters. We also show the dependence of density gradients on laser-plasma interaction and consequently its effect on ion acceleration mechanism. The role of density gradient is of high relevance to investigate, in order to model gas target for ion acceleration experiment. In addition we illustrate the quasi-monoeenrgtic ring shapes electron beam which is a signature effect of axial magnetic field due to the radial spatial field dependence of focused circularly polarized laser.

**Results**

Considering the interaction of a circularly polarised laser pulse propagating in the near critical density plasma in the Y-direction. The electric field of laser can be expressed as $E_L(r,y,t) = (1/2)E_0(r,y,t)\,(e_x + ie_z)exp(-i\omega t + iky) + c.c.$ The laser pulse propagates with the group velocity, $v_{gr} = c^2 k/\omega$ where $k = 2\pi/\lambda$ is the wave vector given by the dispersion relation $c^2 k^2 = \omega^2 - \omega_p^2$. The spatio-temporal intensity profile of laser is considered as Gaussian distribution and can be written as $I = I_0 \exp(-r^2/r_0^2)\exp(-t^2/\tau_0^2)$ where $r_0$ is the transverse radius of laser beam and $\tau_0$ is laser pulse duration. For optimum acceleration in case of MVA[25], the laser spot size should match the size of the self-focusing channel in order to avoid filamentation. The laser beam radius in an underdense plasma can be written as $r = r_0[1 + (x^2/x_R^2)(1-(P/P_c))]^{1/2}$ with critical power $P_c = 17.5\,(n_c/n_e)\,GW$ and the Rayleigh range ($x_R$).

To observe the efficient ion acceleration, the laser pulse energy should be depleted near the target rear side to create the dipole vortex at the exit side of the channel and it should not be wasted to transmission. Thus, the effectiveness of MVA mechanism requires the efficient transfer of laser energy to the fast electrons in the plasma, which are accelerated in the plasma channel along the laser propagation direction. The optimum plasma length[25] $L_{ch} = a_L c\tau (n_c/n_e) K$ can be estimated from the assumption that all laser energy is transferred to the electrons in the plasma channel, where K is the geometry constant (K is 0.1 in 2D case and 0.074 in 3D case). The normalised electric field amplitude of laser is expressed as, $a_L = eE_L/m_e\omega_0 c$. Here e, $m_e$ are fundamental charge and electron mass, $E_L$ is the laser electric field, $\omega_0$ is the laser frequency and c is the speed of light in vacuum.

We show through Fig. (1), dependence of optimum plasma channel length ($L_{ch}$) on relative target density for a given laser field of fixed pulse ($a_L$ = 81, $\tau_0$ = 20 fs, K = 0.074) and defined the three regimes of ion acceleration as;

**I**. when thickness of plasma target is much longer than the optimum plasma channel length ($L_{ch}$) - in such case ion acceleration at rear side of plasma will be due to TNSA mechanism not MVA,

**II**. when thickness of plasma target is smaller than the optimum plasma channel length – the whole pulse energy will not be transmitted to electrons and ions will not accelerate to maximum energy,

**III**. when thickness of plasma target is equal to plasma channel length – in such condition the most of the laser energy will be transferred to electrons and maximum peak ion energy can be achieved. Hence this is treated as the efficient condition for effective ion acceleration via magnetic vortex mechanism.

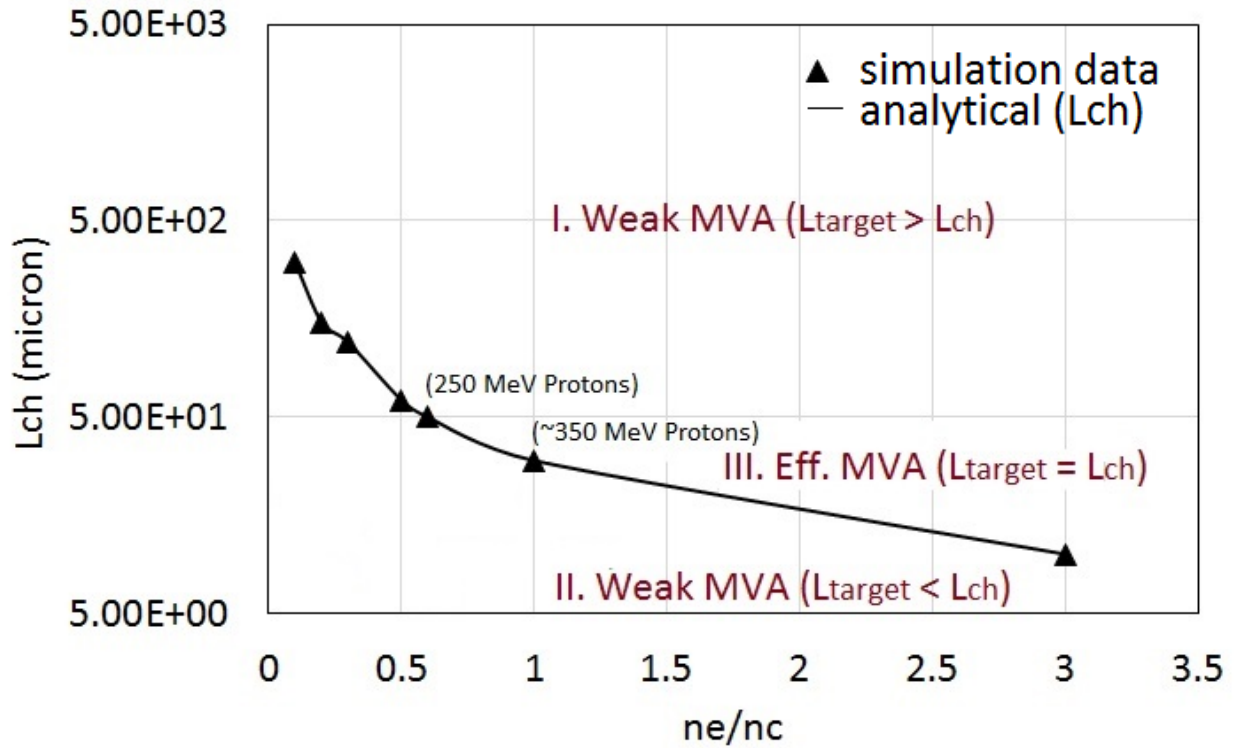

**Figure 1:** Black curve shows the optimum condition for efficient MVA mechanism of ion acceleration however black triangles corresponds to the simulation results.

Fig. 1 shows the optimum condition between the laser-plasma parameters for efficient acceleration of protons to high energy. In Fig. 1 black triangle corresponds to the estimated point (from analytical formulation $L_{ch} = a_L c\tau (n_c/n_e) K$ )); we performed simulation at couple of laser – plasma parameters as indicated by triangles and the maximum energy obtained from simulation results is shown for the points at 0.6 $n_c$ and 1.0 $n_c$. We further explained through Fig. 1, the results of experiment study[27] of gas foil interaction with a 500 mJ, 800nm, 50fs TFL laser system (at NRL) which was performed for plasma densities 0.3 $n_c$ ($L_{ch}$ = 14.9 μm, analytical estimation) and 0.6 $n_c$ ($L_{ch}$ = 5.3 μm, analytical estimation). However in experiment target thickness was considered about 105 μm which is much longer than optimal plasma channel length ($L_{ch}$) for the laser-plasma setup of experiment. As we pointed in Fig. 1 (curve III), that efficient MVA for maximum proton acceleration can be approached when the target thickness is approximately equal to optimum plasma channel length. Thus in order to achieve efficient proton acceleration in experiment, it's important to consider the optimised plasma thickness.

We perform the simulation for the parameters mentioned above 30 μm and demonstrated the evolution of electron (a) and ion (b) charge density (normalized to the critical density) as shown by Fig. 2 at 200 fs, as the laser pulse exits the plasma channel. To illustrate clearly the laser-plasma dynamics the three-dimensional evolution of electron and ion density is shown in Fig 2. The laser pulse propagates through the plasma channel and expels electrons and ions from its path. Thus, the ponderomotive force of the pulse drives electron cavitation. Thus, a spiral channel is formed by the circularly polarized laser pulse due to the interplay between the ponderomotive force and space charge field between electrons and ions (as shown by Fig. 2(a-b)). The plasma electrons trapped and accelerated within these cavitation as shown in Fig. 2(a-b). These fast electrons which are trailing behind the laser pulse formed axial fast electron (as shown in Fig. 3b) which leads the generation of azimuthal magnetic field.

It is evident and shown later that due to radial spatial field dependence of focused circularly polarized laser there will be an axial magnetic field, which in turn leads the generation of $\left(\vec{v}_{x,z} \times \vec{B}_y\right)$ in transverse direction as the transverse component of ponderomotive force. Thus under the resultant force of focused laser pulse there will be ejection of ring of electrons (as shown by Fig. 2a) in laser propagation direction. We show the electron energy distribution (inset of Fig. 2a) of ring shaped electron beam at time instant 200 fs when depleted laser pulse exits the rear side of plasma. Ring shaped high-energy (peaked around 200 MeV) electron beam with quasi-monoenergetic nature may be used for plasma based X-ray source and collimation of positron beam source. The detailed study on generation of specific structured electron beam will be published elsewhere. Since the normalized laser field ($a_0$ = 81) of petawatt laser is much larger than the value of (square root of ration of ion mass to electron mass) $(m_i/m_e)^{1/2}$ , thus the nonlinear dynamics of ion can also be seen in Fig. 2b.

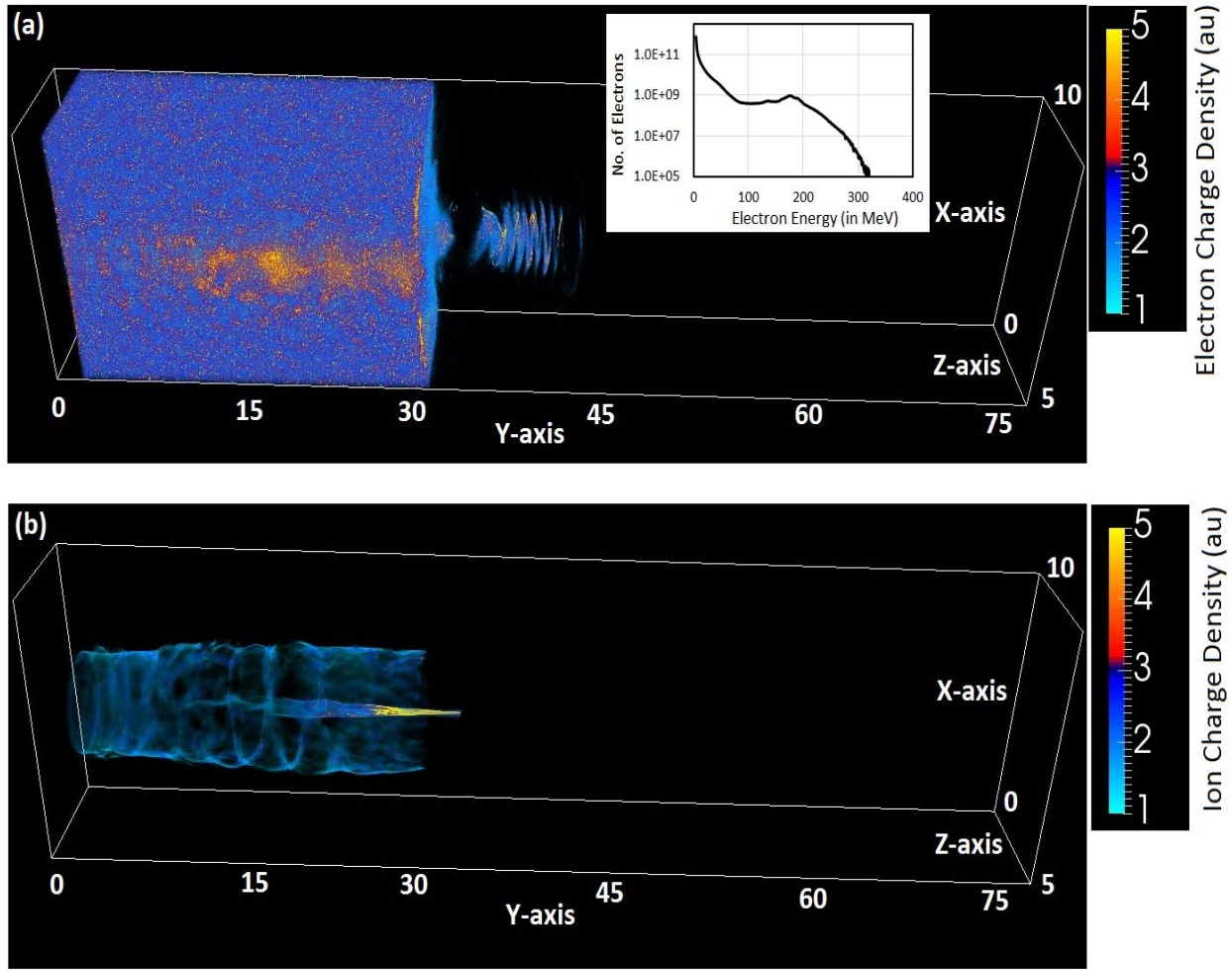

**Figure 2:** Evolution of electron (a) and ion (b) charge density at time instant 200 fs. The laser amplitude, pulse duration, and beam radius are $a_0$ = 81, $\tau_0$ = 20fs and $r_0$=1.5µm respectively. The density of hydrogen plasma target is $n_e$ = $n_c$ and the thickness of plasma target is $L_{target}$ = $L_{ch}$ = 30µm, to utilize the maximum laser energy transfer to plasma electrons. Color bar show the variation in electron and ion charge density. Inset in Fig. 2 (a) shows the electron energy

distribution at time instant 200 fs when depleted laser pulse exits the plasma rear side.

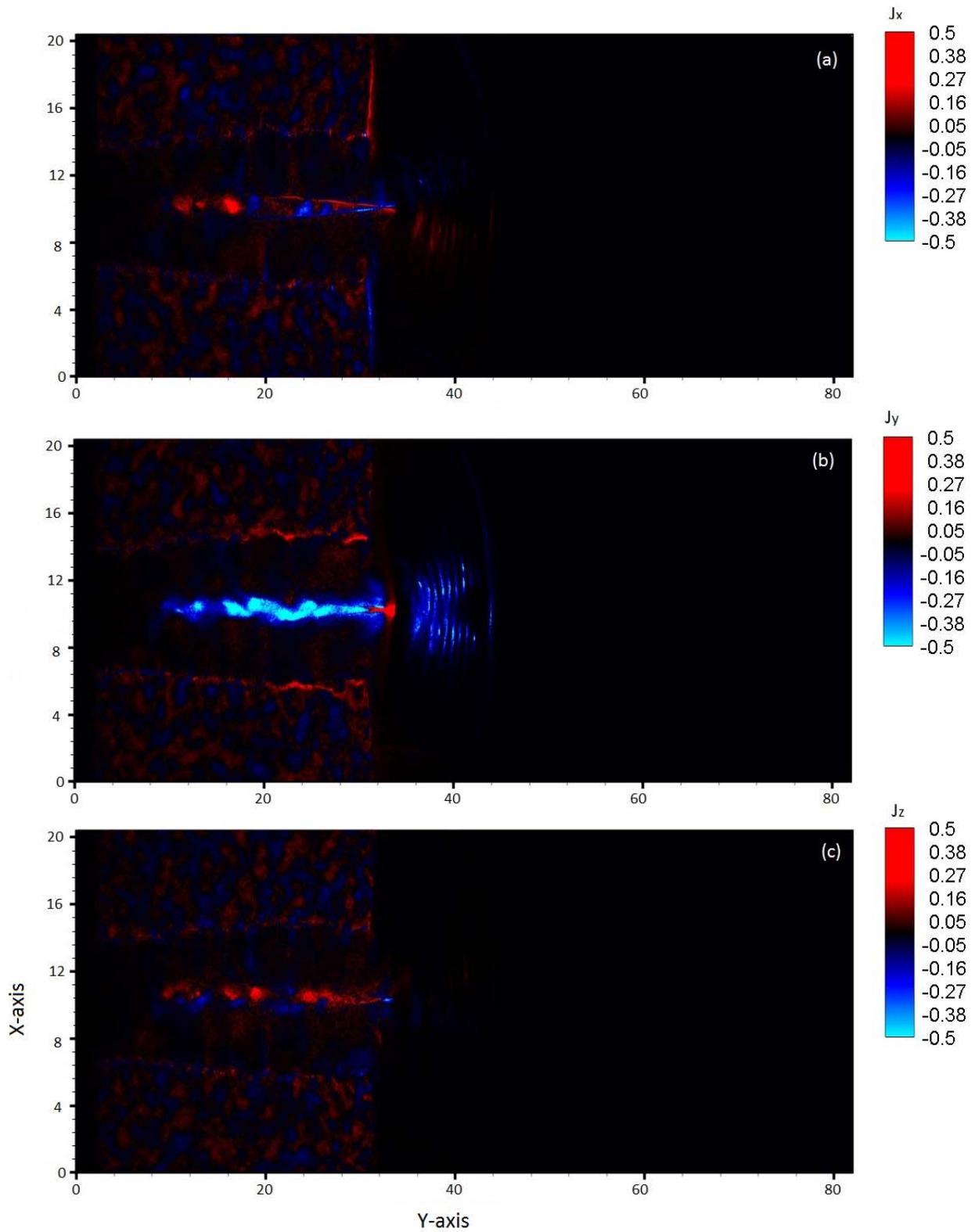

**Figure 3:** The contour plot (a-c) of current density ($J_x$, $J_y$ and $J_z$) (normalised with $n_e\, e\, c$)

along the channel in laser propagation direction at time instant 200 fs. The simulation result shown are for laser –plasma parameters as shown in Figure 2. X-axis and Y-axis are in micron and color bar shows the variation in particle charge density.

As shown in Figure 4 (a), as the laser pulse exits the plasma, the azimuthal magnetic field expands longitudinally and transversally as well. The transverse and longitudinal expansion of azimuthal magnetic field leads to the decrease of electrons drift speed which are coming out from the plasma channel. Thus the slowdown of electron speed increases the density locally and hence the current distribution at rear side of target.

As it is known and can be seen (in Figure 3) that the azimuthal magnetic field is generated by the axial forward current and the return current along the plasma channel wall compensates with the forward current. It can be also seen from Fig. 3 that forward current (blue) is dominating near the axis and the return current (red) dominates away from the axis along the channel wall. Thus, inside the plasma channel forward current dominates over the return current; but at the plasma – vacuum interface since the forward going electrons are forced to move towards target due to magnetic vortex field and there is dominance of return current which decreases the growth of magnetic field. In case of homogenous plasma at plasma – vacuum interface, the abrupt decrease in magnetic field induces a strong electric field, which is responsible to accelerate the ion beam to high energies.

The interaction of focused laser pulse in underdense plasma channel generates the axial and azimuthal magnetic field and can be written as[29]

$$\frac{\partial B_y}{\partial r} = \frac{-\alpha}{1+\alpha^2} n_e \frac{1}{\gamma} \frac{\partial}{\partial r}\left(\frac{<n>}{n_e}\frac{\gamma^2-1}{\gamma}\right) \quad (1)$$

$$\frac{1}{r}\frac{\partial}{\partial r}(rB_Z) = -n_e \frac{<n>}{n_e}\frac{\gamma-1}{\gamma} \quad (2)$$

where $\gamma$ is the relativistic factor, $<n>$ is the modulation in electron density due to the field of laser and can be determined as $<n>/n_e = 1 + (e^2/4 m_e^2 \omega_p^2 \omega^2 \gamma)\nabla^2 \langle \boldsymbol{E} \cdot \boldsymbol{E}^* \rangle$, $B_y$ is the axial magnetic field and $r = \sqrt{x^2 + z^2}$ is the radius of channel in cylindrical coordinate system.

The expression for axial magnetic field ($B_y$) (as given by Eq.1) shows that it can be generated only for CP laser while for LP laser the axial magnetic field is zero. However, the azimuthal magnetic field (given by Eq. 2) generated by the longitudinal current driven by ponderomotive force; can be generated from CP and LP laser both. We confirm the analytical conclusion through simulation results on existence of axial and azimuthal magnetic field from circularly polarised laser pulse in plasmas and shown via Fig. 4. The simulation results show that generated azimuthal magnetic field and axial magnetic field for the laser-plasma parameters adopted here is of the order of $10^5$ T (as shown by Fig. 4a). The simulation results for $B_z$ (~$10^4$ T) of Bulanov[26] is also consistent with the simulation results shown here. Fig. 4 b shows the temporal evolution of axial magnetic field and azimuthal magnetic field which is obtained through the simulation results.

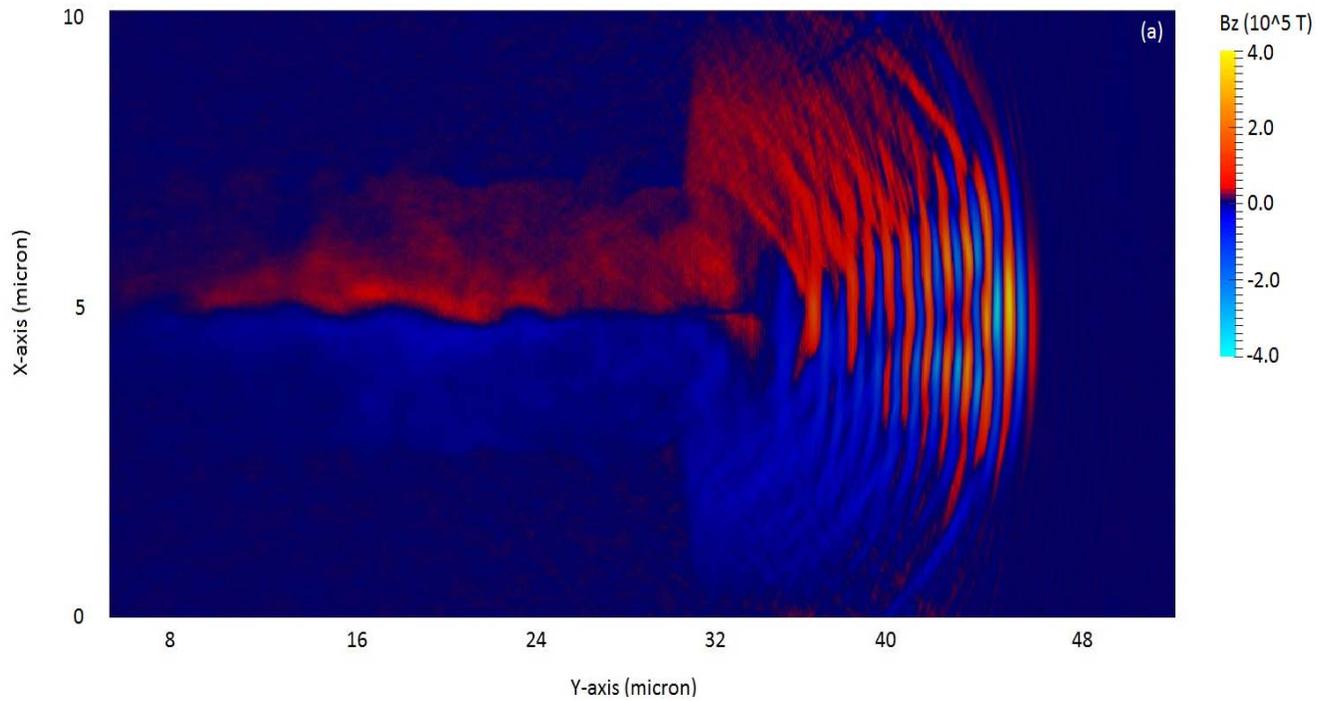

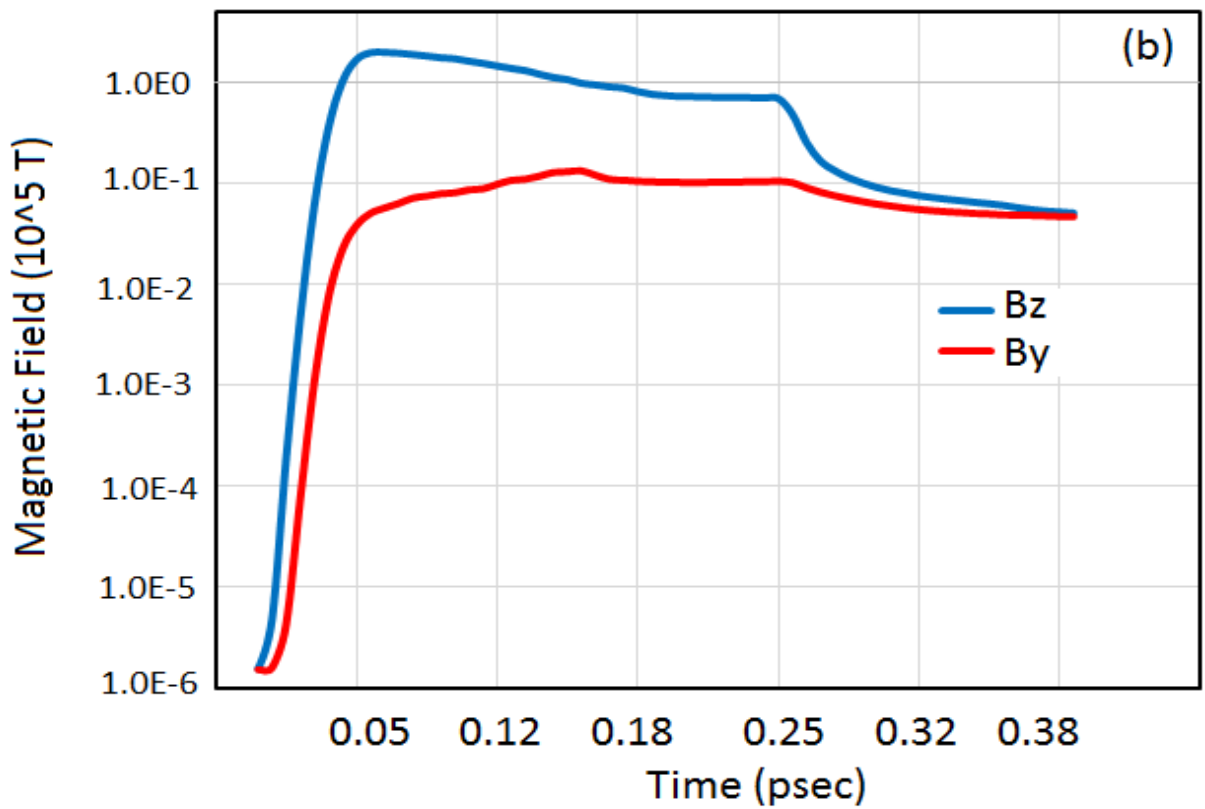

**Figure 4:** Spatial (a) and temporal (b) evolution of self – generated plasma magnetic field from the simulation after the laser exits the plasma at 200fs. The simulation result shown are for laser –plasma parameters as shown in Figure 2.

In our study the transverse and longitudinal expansion of azimuthal magnetic field (shown by Fig. 4a) is playing key role at the rear end of plasma target; to determine the accelerating field (shown by Fig. 5 ) and hence acceleration of ions/ deceleration of electrons. We show in Figure 5, the spatial distribution of electric field and spatial variation of transverse focusing field and (c) longitudinal accelerating field at time instant 200 fs which is responsible for ion beam focusing towards the axis and acceleration along the laser propagation direction.

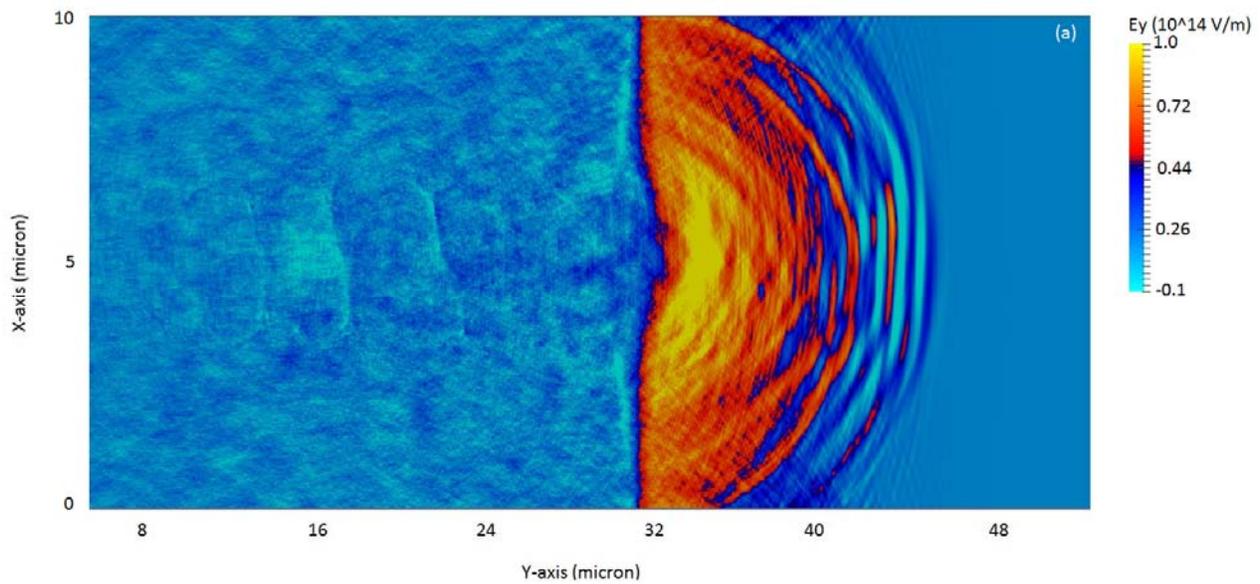

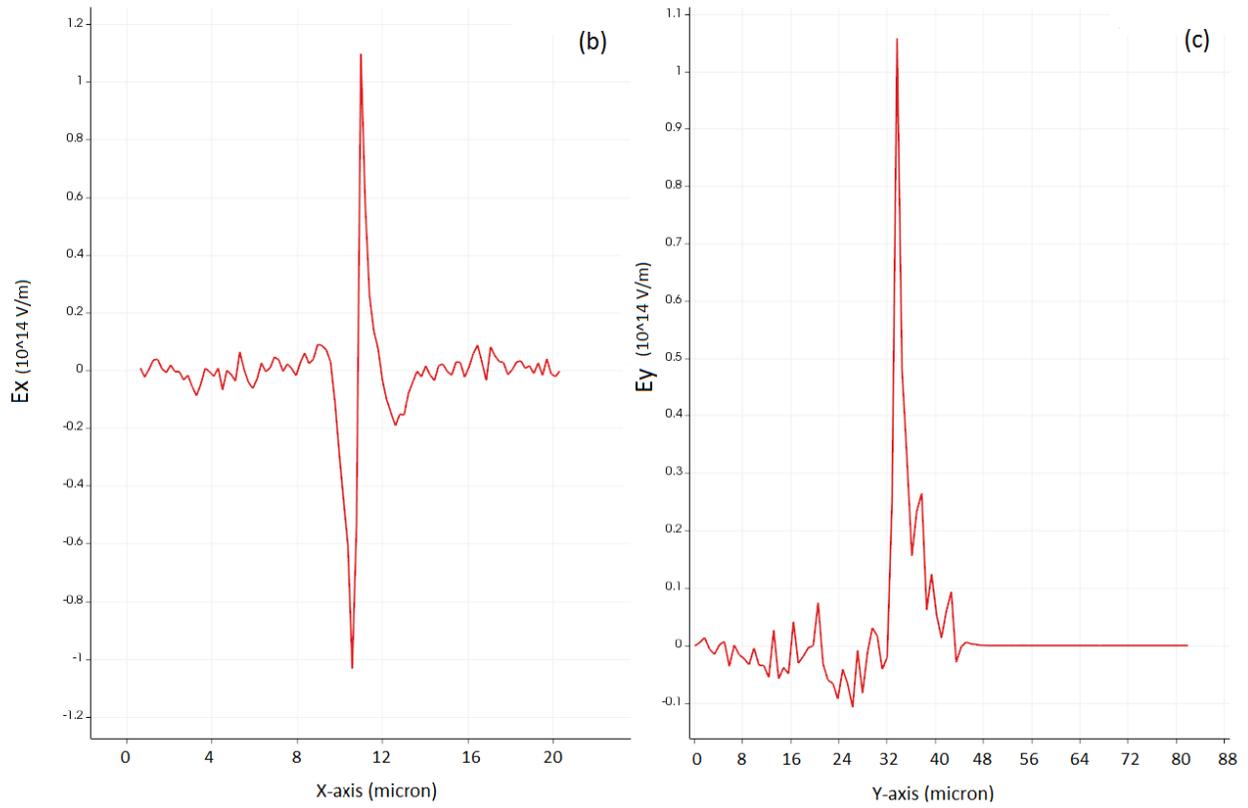

**Figure 5.** The spatial evolution of self – generated plasma electric field (a) and evolution of transverse focusing field normal to the laser propagation direction (b) and along the laser propagation direction (c); after the laser exits the plasma at 200 fs.

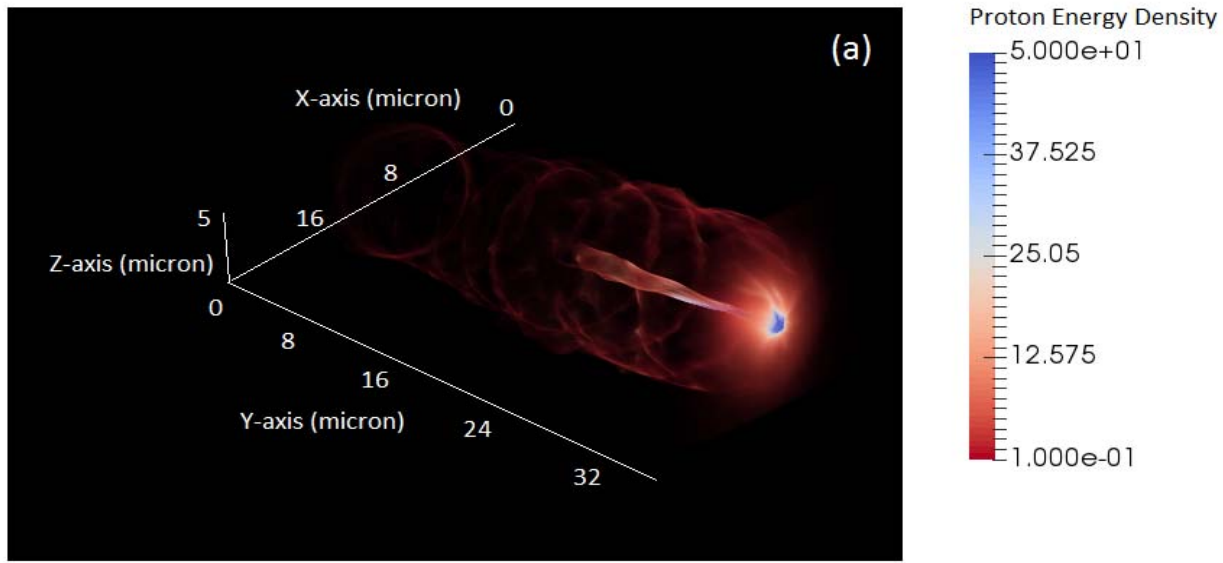

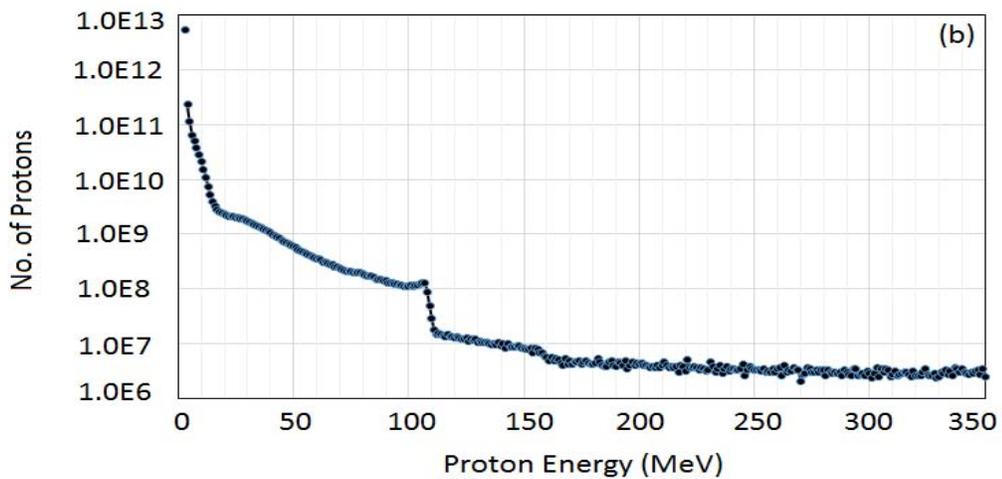

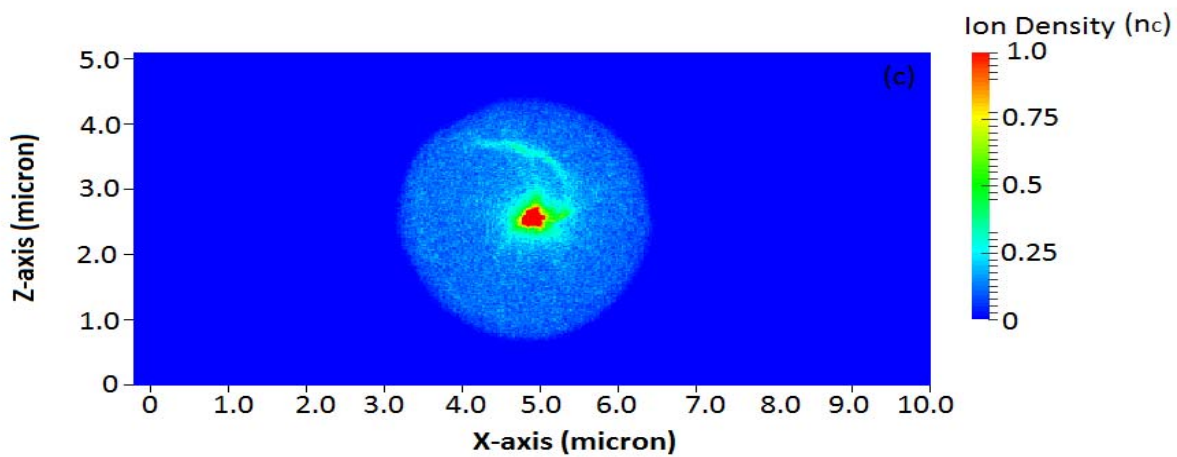

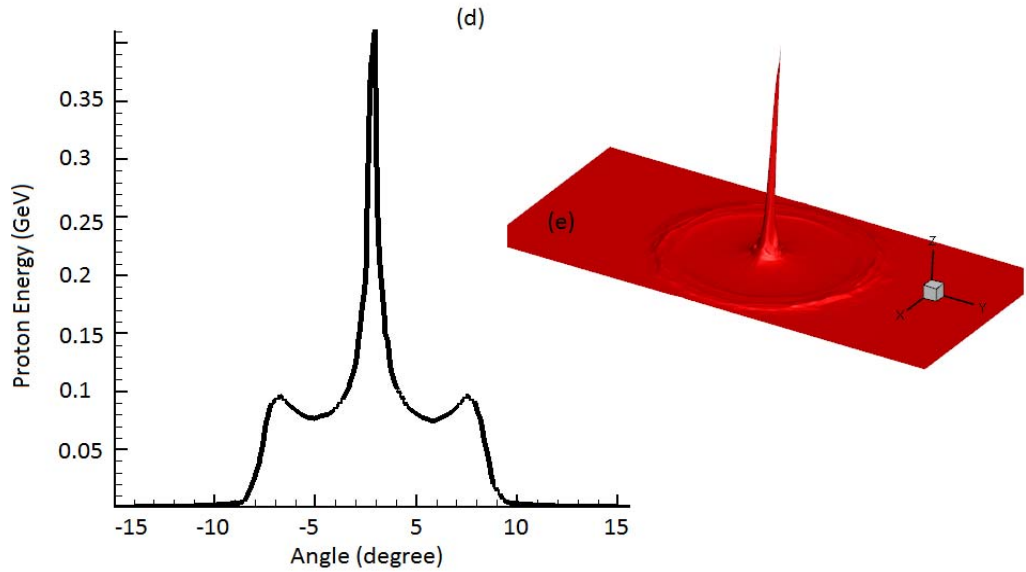

**Figure 6:** (a) The distribution of proton energy density, normalised with $n_c m_e c^2$. Energy (b) and density (c) distribution of protons at time instant 200 fs. The simulation result shown are for laser –plasma parameters as shown in Figure 2. The angular distribution of protons is shown by (d-e).

The generation and acceleration of proton beam from plasma channel is demonstrated in Fig. 6 by the proton energy density distribution at time instant 200 fsec when the laser pulse exits the plasma channel. The distribution of transverse electric field and longitudinal electric field is shown in Fig. 5 (b-c) which are the focusing and accelerating field responsible for focused high energy proton beam. Further the proton beam is characterised (as shown in Figure 6 b-c) by showing the energy and density distribution at time instant of maximum acceleration. Figure 6b shows the energy distribution of protons with maximum proton energy about 350 MeV. Figure 6c shows the spatial density distribution of focused proton beam where high energy protons are concentrated in very small area; which can be attributed to transverse focusing field as shown in Figure 5b. Fig. 6(d-e) shows the angular distribution of proton beam which clearly demonstrates that high energy protons are collimated at smaller divergence angle.

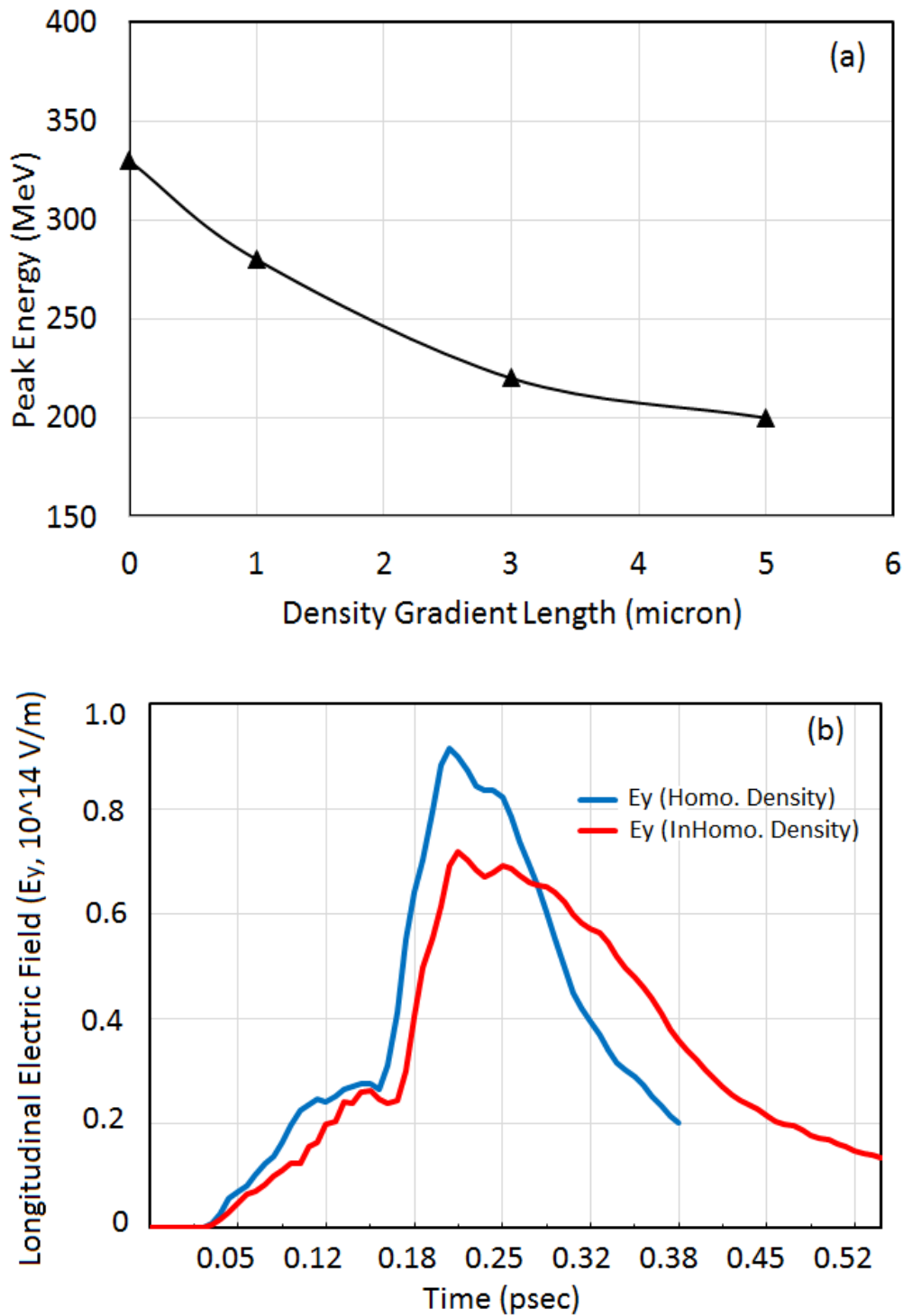

**Figure 7:** Dependence of peak proton energy on density gradient (a) and evolution of longitudinal electric field (b) with time. The longitudinal field evolution explains the dependence of peak proton energy on density gradient. The simulation results are shown at the stage of maximum ion acceleration (at t=200fs). The simulation result shown are for laser –plasma parameters as shown in Figure 2.

The dependence of peak proton energy on density scale length is shown in Figure 7; which is obtained from the simulation results for plasma with density gradient at front and back side of target. The density of flat part of the plasma target is considered as 30 μm while the ramp length is varied 0 (homogenous plasma), 1, 3 and 5 μm to study the role of plasma density gradient at front and rear side of target. The maximum electric field at rear side of target has dependence on density scale length as[9] $E_{max} \propto L^{-1}$, where $L$ is the sheath scale length. In our simulation study, the scale length due to initial plasma density ramp is much larger than the Debye length $\lambda_D = (\varepsilon_0 k_B T_e / n_{e0} e^2)^{1/2}$, and hence its reducing the maximum sheath field at the target rear side and subsequently we obtained the decrease in maximum proton energy.

Since, typical gas-jet targets have linear or parabolic density scale-lengths; too long for the production of quasi-monoenergetic ion beams[31]. Recent experiment[27] demonstrated the 'gas foil' target, available at near critical density at near infrared wavelengths but with the density ramp on both sides. Therefore it's necessary to investigate the role of density ramp in ion acceleration mechanism and to approximate the dependence of peak energy of ions on plasma density gradient.

**Discussions**

The simulation results of proton acceleration with the circularly polarised laser identify the regime of efficient ion acceleration and investigated the dependence of maximum proton energy on laser-plasma parameters. The simulation results also illustrate the experiment results[27] where maximum proton energy can be enhanced by considering the laser-plasma parameter in regime of effective MVA acceleration as shown in Fig.1 by curve III. The experiment study[27] of gas foil interaction with a 500 mJ, 800nm, 50fs TFL laser system (at NRL) is performed for plasma densities 0.3 $n_c$ ($L_{ch}$ = 14.9 µm) and 0.6 $n_c$ ($L_{ch}$ = 5.3 µm); in these two cases the target thickness is about 105 micron (experiment) which is much longer than optimal plasma channel length ($L_{ch}$). As we pointed in Fig. 1 (curve III), that efficient MVA can be approached when the target thickness is equal to optimum plasma channel length. The optimum plasma channel length can be estimated by comparing the total energy of electrons in plasma channel with the pulse energy in self-generated plasma waveguide.

In conclusion, we investigate the regime of efficient MVA which suggests the laser-plasma parameters for experimental illustration in order to achieve high proton energy and higher number of protons by employing near critical density gas target. We numerically studied the mechanism of ion acceleration for circularly polarized laser field in the density window of 0.1 - 1 $n_c$. The proposed density window is preferred in order to get experimental access of near critical density gas targets[27] which may allow to generate the high flux of ions by employment of high rep. rate solid state PW lasers[32]. The interaction of circularly polarised laser field ($a_0$ = 81, $\tau_L$ = 20fs, $r_L$ = 1.5µm) with homogeneous/inhomogeneous plasma shows the proton acceleration for maximum proton energy in range of 200-350 MeV; where we considered the parameter for efficient MVA as scaled by curve III of Figure 1. We also show the dependence of plasma density gradient on peak proton energy since the plasma density profile generated from the gas jet is normally inhomogeneous and the presence of gradients on front and rear side of target affects the generation of fast ions, accelerated from the rear side of target as shown and explained by Figure 7.

Recent experiments[27] demonstrated the ion acceleration with MVA mechanism from gas foil target of near critical density where few MeV proton beam energy was reported. The simulation study presented here suggests to approach efficient proton acceleration in order to achieve high energy in line with experiment results of Helle et al.[27]. We further explored the role of transverse quasi-static field which is responsible for reducing the ion energy spread by focusing the ion beam at plasma-vacuum interface and subsequently focused high energy proton beam where high energy protons are concentrated near the axis of propagation. Finally, the role of polarisation is shown a point of relevance for persistence of focused proton beam along the propagation axis.

In summary, we illustrated here the laser–driven proton acceleration where energetic protons with energies more than 200 MeV and charge in excess of 10 nC[4, 30] and conversion efficiency more than 6% (which implies 2.4 J proton beam out of the 40 J incident laser energy). In addition numerical generation of quasi-monoenergetic ring shaped electron beam is reported first time from circularly polarise laser pulse. These promising ion sources are being developed for applications in nuclear and medical physics, many of which require control of the beam spectral shape, collimation and charge. The underdense and/or near critical gas plasma target can be a better alternative to a thin solid foil because of its low cost, debris-free and can be operated at high repetition rate (>10 Hz) for potential applications. Petawatt laser pulse interaction with matter can produce the conditions that were imagined to occur in astrophysical objects. Thus numerical investigation reported here in regime of ultrashort - ultraintense laser plasma interaction, opens the way for experimental plasma astrophysics to study the property of matter under extreme conditions.

**Method**

To illustrate efficient ion acceleration and underlying microphysics involved in it; we performed a series of 3D PIC simulations using the PIConGPU[33] code for the optimum case. Through the collection of simulation study, we investigate the relevance of circular polarisation, effect of density gradient and influence of self-generated electric and magnetic field on ion beam characteristics. The simulation box size is considered $10\lambda \times 100\lambda \times 5\lambda$ in $x \times y \times z$ directions respectively. A circularly polarized laser pulse ($\lambda = 0.8\mu m$ is the laser wavelength) enters the simulation box from the left boundary along the y direction. The laser field has a Gaussian profile in space and time. The laser amplitude, pulse duration, and beam radius are $a_L = 81$, $\tau_0 = 20fs$ and $r_0 = 1.5\mu m$ respectively. The plasma target proposed here is hydrogen of density $n_i = n_e = 1.0 n_c$ (critical density $n_c = 1.74 \times 10^{21}$ cm$^{-3}$) with variable thickness, to utilize the maximum laser energy transfer to plasma electrons.

**Acknowledgements**

We acknowledge support of the Department of Information Services and Computing, Helmholtz-Zentrum Dresden-Rossendorf (HZDR), Germany; for providing access to the GPU Compute Cluster Hypnos. The authors would also like to thank M. Bussmann and the PIConGPU developer team for fruitful discussions regarding the simulation work.

This work was funded as part of the European Cluster of Advanced Laser Light Sources (EUCALL) project which has received funding from the European Union's Horizon 2020 research and innovation programme under grant agreement No. 654220.

**Authors Contributions Statement**

A.S. conceived the idea, performed simulation and analytical study and completed the article writing based on results.

**Additional Information**

**Competing financial interests:** The author declare no competing financial interests.